\documentclass[11pt,a4paper]{article}

 \usepackage[english]{babel}
 \usepackage[T2A]{fontenc}
 \usepackage[cp1251]{inputenc}
 \usepackage{amsmath}
\usepackage{graphicx}
\usepackage{amssymb}
\usepackage{color}
\usepackage{amsfonts}
\usepackage{wrapfig}
\usepackage{caption}

\newcommand{\be}{\begin{equation}}
\newcommand{\ee}{\end{equation}}

\newcommand{\vk}{\mathbf{k}}
\newcommand{\vl}{\mathbf{l}}
\newcommand{\cB}{{\cal B}}

\begin{document}

\binoppenalty=10000
\relpenalty=10000

\begin{center}
\textbf{\Large{BEHAVIOR OF A BINARY ASYMMETRIC MIXTURE OF INTERACTING PARTICLES IN THE SUPERCRITICAL REGION}}
\end{center}

\vspace{0.3cm}

\begin{center}
M.P.~Kozlovskii and O.A.~Dobush\footnote{e-mail: dobush@icmp.lviv.ua} 
\end{center}

\begin{center}
Institute
for Condensed Matter Physics of the National Academy of Sciences of
Ukraine \\ 1, Svientsitskii Str., 79011 Lviv, Ukraine
\end{center}

 \vspace{0.5cm}

\small We propose a method for describing a phase behavior of a system consisting of particles of two sorts. The interaction of each species is described by interaction potentials containing the repulsive and attractive components. Asymmetry is ensured by different values of the interaction potentials of each sort. The grand partition function of a binary mixture is calculated in the zero-mode approximation A line of critical points, which correspond to different proportions of the components, is calculated for specific values of parameters of the interaction potential. We have obtained an equation that relates the introduced mixing parameter $x$ with the concentration of the fluid. An explicit expression of the pressure of the binary mixture is derived as a function of relative temperature and mixing parameter $x$ to plot the Widom line. It is established that for boundary values of this parameter ($x = 0$ and $x = 1$), the equation of state of a mixture turns into equations of state of its separate species.

\vspace{0.5cm}

PACs: 51.30.+i, 64.60.fd


\normalsize

\section{Introduction} \label{sect_intro}

A crucial part of the theory of phase transitions in multi-particle systems is the elaboration of functional methods. The study of microscopic mechanisms leading to the diversity of phase behavior in mixtures is relevant for prognosticating phase diagrams of systems containing particles of different species. The results obtained by both experimental and theoretical methods are essential in this regard. Among the former are the works \cite{ref1,ref2,ref3} by L.A. Bulavin. We dedicate the present article to the occasion of his 75th birthday.

Theoretical approaches include publications \cite{ref4,ref5}, where the description of multi-component systems is carried out within the mean-field formalism, and also \cite{ref6}, where the study of a binary system is performed in a more general case. The compilation \cite{ref7,ref8} proposes an original approach to explaining phase transitions and critical phenomena, which provides an exact functional representation of the grand partition function of a multi-component model in the method of collective variables \cite{ref9} with the reference system. There a hard-sphere system is the reference system, and the interaction potential contains repulsive and attractive parts. The obtained results were generalized to the case of the Coulomb interaction (RPM) model \cite{ref10}. In \cite{ref12,ref13}, the behavior of a binary symmetric mixture was investigated in the framework of that approach near the critical point using non-Gaussian distributions.

More than 60 years ago, experiments on single fluids identified distinct liquid-like and gas-like structures under supercritical conditions~\cite{refscf1}. Since then, the interest of the scientific community in the comprehensive study~\cite{refscf2,refscf3,refscf4} of supercritical fluids for their widespread technological applications~\cite{refscf5,refscf6} has been steadily increasing due to their high density, solubility, and transport properties. The line between gas-like and liquid-like structure \--- the Widom line~\cite{refscf7} \--- is an extension to the coexistence curve in the supercritical region, which is characterized by maxima in the thermodynamic response functions. Over the past decade this crossover line and the properties of supercritical single fluids has been actively investigated~\cite{refscf8,refscf9}. Recently, M.~Raju et al. have presented an evidence for the existence of Widom lines in binary mixtures from molecular dynamics simulations~\cite{refscf10}. 

The aim of the present article is a description of the phase behavior of a binary fluid in the supercritical region using the cell fluid model \cite{ref14,ref15}. In contrast to the functional approaches mentioned above, we use a ``soft repulsion'' as a reference system. This move provides calculating the grand partition function of the model within a single approach of the collective variables method. This work summarizes the results of \cite{ref16} in the case of an asymmetric binary system. We have found that the critical temperature of the mixture is a function of its composition, described a method for calculating the equation of state of the mixture, and shown its pressure behavior along the Widom line.

\section{Calculating the grand partition function of a binary asymmetric system. } \label{sect_GPF}

We use the results of \cite{ref16} to describe the properties of a binary system of particles. Consider the volume $V$ contains $N_a$ particles of species $a$ and $N_b$ particles of species $b$. The interaction potentials between particles of each sort are modeled by the following type of potential
\begin{equation}\label{2d1}
U_{\delta}(r)  = C_H^{(\delta)} \left\{ A_{\delta} e^{- n_0^{(\delta)} (r_{\delta} - R_0^{(\delta)} )/\alpha_{\delta}} + e^{-\gamma_{\delta} ( r_{\delta} - R_0^{(\delta)} )/\alpha_{\delta}} - 2 e^{ - ( r_{\delta} - R_0^{(\delta)} )/\alpha_{\delta}}\right\}.
\end{equation}
In the latter expression the index $\delta = a,b$ labels the particle species. For the normalization constants
\begin{equation}\label{2d2}
C_H^{(\delta)} = D \frac{ n_0^{(\delta)} }{ n_0^{(\delta)} + \gamma_{\delta} - 2 }; \quad A_{\delta} = \frac{ 2 - \gamma_{\delta} }{ n_0^{(\delta)} }
\end{equation}
we have the following conditions: 1) the minimum of $U_{\delta}(r)$ is at $r_{\delta} = R_0^{(\delta)}$, 2) its absolute value equals the energy of dissociation $-D$. For simplicity assume $\alpha_a = \alpha_b = \alpha$ is the effective interaction radius and $R_0^{(a)} = R_0^{(b)} = R_0$. Moreover we use $R_0$-units to measure all the distance quantities. The interaction potentials $U_{a}(r)$ and $U_{b}(r)$ differ from each other by the parameters $\gamma_a \neq \gamma_b$ and $n_0^{(a)} \neq n_0^{(b)}$. The values of these parameters used in the present research are given in Appendix~\ref{appA}. However, they might be changed to reflect other particular systems. The fluid system is henceforth modeled as a cell fluid \cite{ref17}, in which the total volume $V$ of the system is divided into $N_v$ cubic cells each of the side $c$ and the volume $v = c^3$. 
The lattice analog of \eqref{2d1} is expressed as
\begin{equation}\label{2d3}
 U_{l_{12}}^{(\delta)}  = C_H^{(\delta)} \left\{ A_{\delta} e^{ - \dfrac{n_0^{(\delta)} (l_{12} - c )}{\alpha_R c} } + e^{ -\dfrac{\gamma_{\delta} (l_{12} - c)}{\alpha_R c} }  
 - 2 e^{ - \dfrac{l_{12} - c}{\alpha_R c} } \right\}.
\end{equation}
here $\alpha_R = \alpha/R_0 $ is a dimensionless quantity, $l_{12} = |\vl_1 - \vl_2|$ is a distance between the cell vectors $\vl_1, \vl_2 \in \Lambda$:
\begin{equation}\label{2d4}
\Lambda  =  \Big\{  \vl = (l_x, l_y, l_z) | l_i = c \cdot n_i; n_i = 1,2, \ldots, N_i; \; i = x,y,z; N_i = (N_v)^{1/3} \Big\}.
\end{equation}
In the thermodynamic limit $V \rightarrow \infty$, $N_v \rightarrow \infty$, $v = V/N_v = const $. The Fourier transform of \eqref{2d3} has the form
\begin{equation}\label{2d5}
\Phi_\delta (k) = \Psi_\delta (k) + \Phi^{(r)}_\delta (k) - \Phi^{(a)}_\delta (k),
\end{equation}
where both $\Phi^{(r)}_\delta (k) $ and $\Psi_\delta (k)$ are the repulsive parts of the interaction between species $\delta$, and $\Phi^{(a)}_\delta (k) $ is the attraction.
\begin{align}\label{2d6}
& \Psi_\delta (k) = C_H^{(\delta)} \! A_\delta 8 \pi e^{ \frac{n_0^{(\delta)}}{\alpha_R}} \left( \! \frac{\alpha_R}{n_0^{(\delta)}} \! \right)^{\!\! 3} \!\! \left[ 1 +  \left( \frac{\alpha_R}{n_0^{(\delta)}} c k \right)^{\!\! 2}  \right]^{-2} \!\!\!\!, \nonumber \\ 
& \Phi^{(r)}_\delta (k) = C_H^{(\delta)} 8 \pi e^{\gamma_\delta/\alpha_R} \left( \frac{\alpha_R}{\gamma_\delta} \right)^{\!\! 3} \!\! \left[ 1 +  \left( \frac{\alpha_R}{\gamma_\delta} c k \right)^{\!\! 2} \right]^{-2} \!\!\!\!, \nonumber \\
&  \Phi^{(a)}_\delta (k) = C_H^{(\delta)} 16 \pi e^{1/\alpha_R} \alpha_R^3 \left[ 1 +  \left( \alpha_R c k \right)^2  \right]^{-2} . 
\end{align}
Recall that $\delta = a,b$. According to the results of \cite{ref16}, we write the expression for the grand partition function of a binary cell fluid model as follows
	\begin{align}\label{2d7}
	& \Xi = g_V^{(a)} g_V^{(b)} \exp \left[ N_v \left( a_0 - \frac{\beta \tilde \mu_a^2}{2 W_a (0)} \right) \right] \exp \left[ N_v \left( b_0 - \frac{\beta \tilde \mu_b^2}{2 W_b (0)} \right) \right] \times  \\ 
	& \int \! (dt_{\vk}^{(a)})^{N_v} \exp \! \Bigg[ N_v^{\frac12} \! \left( \! \frac{\tilde \mu_a}{W_a (0)} + a_1 \! \right) t_0^{(a)} \! + \frac12 \sum \limits_{\vk \in \cB_c} \! D_a(k) t_{\vk}^{(a)}t_{-\vk}^{(a)} + \! \sum \limits_{n=3}^{\infty} \frac{a_n}{n!} N_v^{\frac{2-n}{n}} \!\!\! \sum_{\substack{\vk_1,...,\vk_n \\ \vk_{i}\in \cB_c}} \!\! t_{\vk_1}^{(a)} \ldots t_{\vk_n}^{(a)} \delta_{\vk_1 + \ldots + \vk_n} \Bigg] \nonumber  \\
	&  \int \! (dt_{\vk}^{(b)})^{N_v} \exp \! \Bigg[ N_v^{\frac12} \! \left( \! \frac{\tilde \mu_b}{W_b (0)} + b_1 \! \right) t_0^{(b)} \! + \frac12 \sum \limits_{\vk \in \cB_c} \! D_b(k) t_{\vk}^{(b)}t_{-\vk}^{(b)} + \! \sum \limits_{n=3}^{\infty} \frac{b_n}{n!} N_v^{\frac{2-n}{n}} \!\!\! \sum_{\substack{\vk_1,...,\vk_n \\ \vk_{i}\in \cB_c}} \!\! t_{\vk_1}^{(b)} \ldots t_{\vk_n}^{(b)} \delta_{\vk_1 + \ldots + \vk_n} \Bigg]  \nonumber. 
	\end{align}
here $\tilde \mu_{\delta} = \mu_\delta - \mu_\delta^* \beta_c^{(\delta)}/\beta$, де $\mu_\delta^*$ is a constant, and $\beta_c^{(\delta)}$ is the inverse critical temperature of species $\delta \; (\delta = a,b)$. The effective potentials of interaction~\cite{ref16} read
\begin{align}\label{2d8}
& W_\delta (k) = \Phi^{(a)}_\delta (k) - \Phi^{(r)}_\delta (k) - \Psi_\delta (k) + \frac{\beta_c^{(\delta)}}{\beta} \Psi_\delta (0), \\
& D_a (k) = a_2 - \frac{1}{\beta W_a(k)}, \quad  D_b (k) = b_2 - \frac{1}{\beta W_b(k)}. \nonumber
\end{align}
\begin{equation}\label{2d9}
g_V^{(\delta)} = \prod \limits_{\vk \in \cB_c} (2 \pi \beta W_\delta (k) )^{-\frac12}
\end{equation}
The cumulants $a_n$ are as follows
\begin{align}\label{2d10}
& a_0 = \ln T_0(\alpha_a^*, p_a), \quad a_1 = \frac{T_1(\alpha_a^*, p_a)}{T_1(\alpha_a^*, p_a)}, \nonumber \\
& a_2 = \frac{T_2(\alpha_a^*, p_a)}{T_1(\alpha_a^*, p_a)} - a_1^2, \\
& a_3 = \frac{T_3(\alpha_a^*, p_a)}{T_1(\alpha_a^*, p_a)} - a_1^3 + 3 a_1 a_2, \nonumber \\
& a_4 = \frac{T_4(\alpha_a^*, p_a)}{T_1(\alpha_a^*, p_a)} - a_1^4 - 6 a_1^2 a_2 - 4 a_1 a_3 - 3 a_2^2,  \nonumber
\end{align}
where the special functions $ T_n(\alpha_a^*, p_a)$ read
\begin{equation}\label{2d11}
T_n(\alpha_a^*, p_a) = \sum \limits_{m=0}^{\infty} \frac{(\alpha_a^*)^m}{m!} m^n e^{-p_a m^2}.
\end{equation}
Here $\alpha_a^* = v \exp[\beta_c^{(a)}\mu_a^*]$, and the parameter $p_a$ is expressed via the quantity $\Psi_a (0)$:
\begin{equation}\label{2d12}
p_a = \frac{1}{2} \beta_c^{(a)} \Psi_a (0), 
\end{equation}
so it depends on the parameters $n_0^{(a)}$ and $\gamma_a $ of the interaction potential $U_{l_{12}}^{(a)}$ given by \eqref{2d3}.

The cumulants $b_n$ are expressed by the formulas similar to \eqref{2d9}, where the special functions $S_n(\alpha_b^*, p_b)$ 
\begin{equation}\label{2d13}
S_n(\alpha_b^*, p_b) = \sum \limits_{m=0}^{\infty} \frac{(\alpha_b^*)^m}{m!} m^n e^{-p_b m^2}
\end{equation}
are substituted for the functions $T_n(\alpha_a^*, p_a)$, $\alpha_b^* = v \exp[\beta_c^{(b)}\mu_b^*]$, and the $p_b$ is as follows
\begin{equation}\label{2d14}
p_b = \frac{1}{2} \beta_c^{(b)} \Psi_b (0). 
\end{equation}

At the present stage the expression \eqref{2d7} of the grand partition function does not contain any approximations, but has infinite series in the exponent. To deal with this case, we use the traditional approximation for the phase transition theory, assuming $ a_n $ and $ b_n $  with $ n \geq 5 $ are zero. In general, the calculation scheme below allows us to take into account the higher powers of the variable $\rho_{\vk}^{(\delta)}$, as we did in \cite{ref18} when investigating the second-order phase transition for the Ising model.

So we restrict the problem to the approximation of the $\rho^4$-model, make the change of variables to eliminate the third power of variables $\rho_{\vk}^{(\delta)}$ and read
\begin{equation}\label{2d15}
\Xi =  g_v^{(a)} g_v^{(b)} \int (d \rho^{(a)})^{N_v} e^{N_v E_\mu^{(a)}} \exp[E(M_a,\rho_{\vk}^{(a)} )]  \int (d \rho^{(b)})^{N_v} e^{N_v E_\mu^{(b)}} \exp[E(M_b,\rho_{\vk}^{(b)} )].
\end{equation}
For $E_\mu^{(\delta)}$ we have
\begin{align}\label{2d16}
&E_\mu^{(a)} = a_0 - \frac{\beta \tilde \mu_a^2}{2 W_a(0)} + m_a \left( a_1 + \frac{\tilde \mu_a}{W_a(0)} \right)  + \frac12 m_a^2 d_a(0) + \frac{a_3^4}{8 a_4^3} ,  \\
& E_\mu^{(b)} = b_0 - \frac{\beta \tilde \mu_b^2}{2 W_b(0)} + m_b \left( b_1 + \frac{\tilde \mu_b}{W_b(0)} \right)  + \frac12 m_b^2 d_b(0) + \frac{b_3^4}{8 b_4^3},  \nonumber 
\end{align}
\begin{align*}
& \frac{\tilde \mu_a}{W_a(0)} = M_a - a_1 - m_a d_a(0) + \frac{a_3^3}{a_4^2}, \nonumber \\
& \frac{\tilde \mu_b}{W_b(0)} = M_b - b_1 - m_b d_b(0) + \frac{b_3^3}{b_4^2}, \nonumber
\end{align*}
here $m_a = - \dfrac{a_3}{a_4}$, $m_b = - \dfrac{b_3}{b_4}$. The functions $E(M_\delta,\rho_{\vk}^{(\delta)} )$ are given by
\begin{align}\label{2d17}
& E(M_a,\rho_{\vk}^{(a)} )  = N_v^{\frac12} M_a \rho_0^{(a)} + \frac12 \sum \limits_{\vk \in \cB_c} d_a (k) \rho_{\vk}^{(a)} \rho_{-\vk}^{(a)} + \frac{a_4}{24} \frac{1}{N_v} \sum_{\substack{\vk_1,...,\vk_4 \\ \vk_{i}\in \cB_c}} \rho_{\vk_1}^{(a)} \ldots \rho_{\vk_4}^{(a)} \delta_{\vk_1 \ldots \vk_4},   \nonumber \\
& E(M_b,\rho_{\vk}^{(b)} )  =  N_v^{\frac12} M_b \rho_0^{(b)} + \frac12 \sum \limits_{\vk \in \cB_c} d_b (k) \rho_{\vk}^{(b)} \rho_{-\vk}^{(b)}  + \frac{b_4}{24} \frac{1}{N_v} \sum_{\substack{\vk_1,...,\vk_4 \\ \vk_{i}\in \cB_c}} \rho_{\vk_1}^{(b)} \ldots \rho_{\vk_4}^{(b)} \delta_{\vk_1 \ldots \vk_4}. 
\end{align}
The quantities $d_\delta (k)$ are expressed by
\begin{align}\label{2d18}
& d_a (k) = \tilde a_2 - \frac{1}{\beta W_a(k)}, \quad \tilde a_2 = a_2 - \frac{a_3^2}{2 a_4},  \\
& d_b (k) = \tilde b_2 - \frac{1}{\beta W_b(k)}, \quad \tilde b_2 = b_2 - \frac{b_3^2}{2 b_4}. \nonumber
\end{align}
According to \eqref{2d6}, we read
\begin{align}\label{2d19}
& \Phi_\delta^{(a)} (0) = B_\delta \Phi_\delta^{(r)} (0), \nonumber \\
& \Psi_\delta^{(a)} (0) = A_\gamma^{(\delta)} \Phi_\delta^{(r)} (0),\\
&  B_\delta = 2 \gamma_\delta^3 \exp\left[\frac{1-\gamma_\delta}{\alpha_R}\right], \nonumber \\
&  A_\gamma^{(\delta)}  = A_\delta \left( \frac{\gamma_\delta}{n_0^{(\delta)}}\right)^3  \exp\left[\frac{n_0^{(\delta)}-\gamma_\delta}{\alpha_R}\right]. \nonumber
\end{align}
So we can rewrite $p_a$ and $p_b$ given by \eqref{2d12} and \eqref{2d14} as follows
\begin{align}\label{2d20}
& p_a \! = \! \frac12 \beta_c^{(a)} \Phi_a^{(r)}(0) A_\gamma^{(a)}, \quad p_b \! = \!  \frac12 \beta_c^{(b)} \Phi_b^{(r)}(0) A_\gamma^{(b)}, 
\end{align}
and also
\begin{align}\label{2d21}
& W_a(0) = \Phi_a^{(r)}(0) [B_a - 1 + \tau_a A_\gamma^{(a)}], \\
& W_b(0) = \Phi_b^{(r)}(0) [B_b - 1 + \tau_b A_\gamma^{(b)}].\nonumber
\end{align}
The quantites $\tau_a$ and $\tau_b$ are the relative temperatures of pure components of the mixture (separately for species $a$ and $b$, respectively)
\begin{equation}\label{2d22}
\tau_a = \frac{T-T_c^{(a)}}{T_c^{(a)}}, \quad \tau_b = \frac{T-T_c^{(b)}}{T_c^{(b)}}.
\end{equation}
Obviously, the critical temperature of the mixture $T_c$ is different from $T_c^{(\delta)}$. Taking that into account write the following
\begin{align}\label{2d23}
& \tau_a = \frac{\tau T_c}{T_c^{(a)}} + \frac{T_c-T_c^{(a)}}{T_c^{(a)}}, \\
& \tau_b = \frac{\tau T_c}{T_c^{(b)}} + \frac{T_c-T_c^{(b)}}{T_c^{(b)}},\nonumber
\end{align}
where $\tau$ is the relative temperature of the mixture.
\begin{equation}\label{2d24}
\tau = \frac{T - T_c}{T_c}.
\end{equation}
From \eqref{2d22} and \eqref{2d24} we find that
\begin{align}\label{2d25}
& W_{ \! a} \! (0) \!  = \!  \Phi_a^{( \! r \! )} \! (0) \!\! \left[ \! B_a \! - \! 1 \! + \! \tau A_\gamma^{( \! a \! )} \!  \frac{ T_c}{T_c^{( \! a \! )}} \! + \! \frac{T_c \! - \! T_c^{( \! a \! )}}{T_c^{( \! a \! )}} A_\gamma^{( \! a \! )} \! \right] \!\! , \nonumber \\
& W_{ \! b} \! (0)  \! = \!  \Phi_b^{( \! r \! )} \! (0) \!\! \left[ \! B_b \! - \! 1 \! + \! \tau A_\gamma^{( \! b \! )} \!  \frac{ T_c}{T_c^{( \! b \! )}} \! + \! \frac{T_c \! - \! T_c^{( \! b \! )}}{T_c^{( \! b \! )}} A_\gamma^{( \! b \! )} \! \right] \!\! .
\end{align}
In the boundary case, either $T_c = T_c^{(a)}$ or $T_c = T_c^{(b)}$, the expressions given by \eqref{2d25} transform into \eqref{2d21}. The quantities $T_c^{(\delta)}$ are defined from the conditions $d_\delta (0) = 0$ and are equal to
\begin{align}\label{2d26}
& k_B T_c^{(a)} = \tilde a_2 W_a (0,T_c^{(a)}) = \tilde a_2 \Phi_a^{(r)}(0) (B_a -1),\nonumber \\
& k_B T_c^{(b)} = \tilde b_2 W_b (0,T_c^{(b)}) = \tilde b_2 \Phi_b^{(r)}(0) (B_b -1).
\end{align}
Note that because of \eqref{2d21} we use the following equalities
\begin{align}\label{2d27}
& \Phi_a^{(r)} (0) A_\gamma^{(a)} = 2 p_a k_B T_c^{(a)}, \nonumber \\
& \Phi_b^{(r)}(0) A_\gamma^{(b)} = 2 p_b k_B T_c^{(b)},
\end{align}
to find the critical temperature of the mixture.

\section{The zero-mode approximation}\label{sect_mfa}

Recently in \cite{ref16}, we have shown that the zero-mode approximation describes well the phase behavior of a one-component fluid. It is a mean-field type approximation taking into account the contribution from the variable $\rho_{\vk}$ at $\vk =0$. The contributions from $\rho_{\vk}$ at $\vk \neq 0$ are important near the critical point. Let's investigate the role of $\rho_0^{(\delta)}$ in the formation of a phase diagram of the mixture.  

The grand partition function of the mixture in the zero-mode approximation is as follows \begin{equation}\label{3d1}
\Xi_0 = g_V^{(a)} g_V^{(b)} \exp \big[N_v \big( E_\mu^{(a)} + E_\mu^{(b)} + E_a (\bar \rho_a) + E_b (\bar \rho_b) \big) \big]. 
\end{equation} 
The expressions of $E_\mu^{(\delta)}$ are given in \eqref{2d16}, and for $E_0 (\bar \rho_\delta)$ read
\begin{align}\label{3d2}
& E_a (\bar \rho_a) = M_a \bar \rho_a + \frac12 d_a(0) \bar\rho_a^2 + \frac{a_4}{24}\bar \rho_a^4,\nonumber \\
& E_b (\bar \rho_b) = M_b \bar \rho_b + \frac12 d_b(0) \bar\rho_b^2 + \frac{b_4}{24}\bar \rho_b^4.
\end{align}
It is convenient to use the Laplace method~\cite{ref19} to calculate \eqref{3d1}. Thereby the values $\bar \rho_a$ and $\bar \rho_b$ correspond to maxima of the function $E_0 (\bar \rho_a)$ та $E_0 (\bar \rho_b)$, respectively. So the following equations determine $\bar \rho_a$ and $\bar \rho_b$ 
\begin{align}\label{3d3}
& M_a + d_a(0)\bar \rho_a + \frac{a_4}{6}\bar \rho_a^3 = 0, \nonumber \\
& M_b + d_b(0)\bar \rho_b + \frac{b_4}{6}\bar \rho_b^3 = 0.
\end{align}
The equalities \eqref{2d27} define the critical temperatures of each subsystem, which are also calculated in~\cite{ref16}. The expression \eqref{3d1} contains independent contributions from the particle of different species. 

Now introduce the mixing parameter $x$ as follows
\begin{equation}\label{3d4}
\bar \rho_a = x \rho_+, \quad \bar \rho_b = (1 - x) \rho_+.
\end{equation}
Easy to see that 
\begin{align}\label{3d5}
& \rho_+ = \bar \rho_a + \bar \rho_b,  \\
& x = \frac{\bar \rho_a}{\bar \rho_a + \bar \rho_b}, \quad 1-x = \frac{\bar \rho_b}{\bar \rho_a + \bar \rho_b} .\nonumber
\end{align}
The mixing parameter $x$ is like the concentration in different space. In Section~\ref{sect_density} we show the relation between the parameter $x$ and the concentration of a solution. In terms of $\rho_+$ and $x$ the expressions \eqref{3d2} have the form
\begin{align}\label{3d6}
& E_a (x, \rho_+)  =  M_a x \rho_+ + \frac12 d_a(0) x^2 \rho_+^2 + \frac{a_4}{24} x^4 \rho_+^4,\nonumber \\
& E_b (x, \rho_+) =  M_b (1 - x) \rho_+ + \frac12 d_b (0) (1 - x)^2 \rho_+^2  + \frac{b_4}{24} (1 - x)^4 \rho_+^4.
\end{align}
Their sum (see \eqref{3d1}) is as follows
\begin{equation}\label{3d7}
E_{ab}(x, \rho_+) = M_{ab} + \frac{1}{2}D_{ab} \rho_+^2 - a_+ \rho_+^4,
\end{equation}
where we use denotations
\begin{align}\label{3d8}
& M_{ab} = x M_a + (1-x) M_b , \nonumber \\
& D_{ab} = x^2 d_a(0) + (1-x)^2 d_b(0),  \\
& a_+ = - x^4 a_4 - (1 - x)^4 b_4 . \nonumber
\end{align}
In terms of $\rho_+$ and $x$ the grand potential of the mixture read
\begin{equation}\label{3d9}
\Omega = -k_B T \ln \Xi_0 (x, \rho_+).
\end{equation}
In order to minimize it with respect to $\rho_+$ we have to solve the following equation
\begin{equation}\label{3d10}
\frac{\partial E_{ab} (x, \rho_+)}{\partial \rho_+} = 0,
\end{equation}
namely
\begin{equation}\label{3d11}
M_{ab} = - D_{ab}(x,\tau) \rho_+ + \frac{a_+}{6}\rho_+^3.
\end{equation}
An equation of this type also holds for a single system. The equation \eqref{3d11} is peculiar for the coefficients $D_{ab} (x,T)$ and $a_+(x)$ which are functions of the mixing parameter $x$. Moreover, $D_{ab}$ also depends on the relative temperature of the mixture $\tau$, while the relative temperatures $\tau_a$ and $\tau_b$ characterize pure species $a$ and $b$, respectively.

The effective chemical potential of the mixture $M_{ab}$ is a variable of the same type as either $M_a$ for the subsystem of particles of species $a$, or $M_b$ for species $b$. Each of them associate with $\mu_a$ and $\mu_b$. Using the quantity $M_{ab}$ (or $M_a$, $M_b$) is convenient because the first order phase transition occurs at $M_a = 0$ for the subsystem of particles of species $a$, $M_b = 0$  for the subsystem  of particles of species $b$ and $M_{ab} = 0$ for the mixture of particles of both species $a$ and $b$.  

\begin{figure}[h!]
	\centering \includegraphics[width=0.5\textwidth]{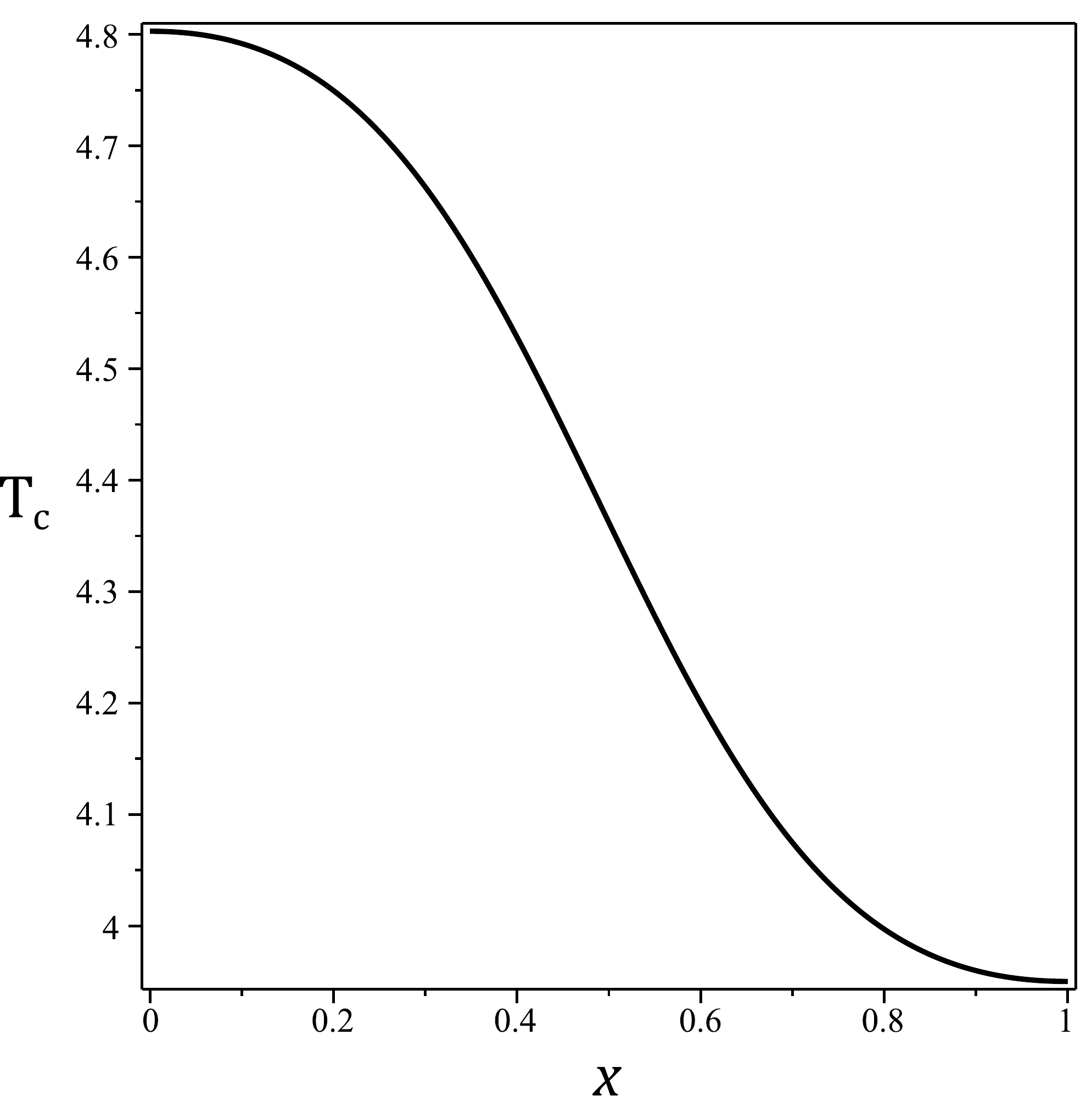}\\
	\vskip-3mm\caption{Plot of the line of critical temperatures of the binary mixture.}\label{fig1}
\end{figure}

Both the equations \eqref{3d11} and \eqref{3d3} are of the same type, therefore, the condition 
\begin{equation}\label{3d12}
D_{ab} (x, T_c) = 0.
\end{equation} 
defines the critical temperature of the liquid-gas phase transition of a binary system. Taking into account \eqref{3d8} we obtain the equation for $T_c = T_c(x)$
\begin{equation}\label{3d13}
x^2 \!\! \left[\tilde a_2 - \frac{1}{\beta_c W_a(0,T_c)}\right] + (1-x)^2 \!\! \left[\tilde b_2 - \frac{1}{\beta_c W_b(0,T_c)}\right] \! = \! 0.
\end{equation}
Use \eqref{2d26} and \eqref{2d27} to solve the latter equation and get
\begin{align}\label{3d14}
& \tilde a_2 W_a(0,T_c) = T_{ca} + 2 \tilde a_2 p_a T_c, \nonumber \\
& \tilde b_2 W_b(0,T_c) = T_{cb} + 2 \tilde b_2 p_b T_c,
\end{align}
where 
\begin{align}\label{3d15}
& T_{ca} = T_c^{(a)}( 1 - 2 \tilde a_2 p_a), \nonumber \\
& T_{cb} = T_c^{(b)}( 1 - 2 \tilde b_2 p_b).
\end{align} 
Then rewrite \eqref{3d13} as follows
\begin{align}\label{3d16}
&\ae \tilde a_2 W_a(0,T_c) \tilde b_2 W_b(0,T_c) - x^2 T_c \tilde b_2 W_b (0,T_c)/\tilde b_2  - (1-x)^2 T_c \tilde a_2 W_a (0,T_c)/\tilde a_2 = 0
\end{align}
Here the expression \eqref{3d13} is multiplied by $W_a(0,T_c) W_b(0,T_c) \neq 0$, and for $\ae$ read
\begin{equation}\label{3d17}
\ae = \frac{x^2}{\tilde b_2} + \frac{(1-x)^2}{\tilde a_2}. 
\end{equation}
Substituting the expressions \eqref{3d14} in the formula \eqref{3d16} we obtain the quadratic equation
\begin{equation}\label{3d18}
f_2 T_c^2 + f_1 T_c + f_0 = 0,
\end{equation}
where
\begin{align}\label{3d19}
& f_2 = 4 \ae p_a p_b \tilde a_2 \tilde b_2 - 2 x^2 p_b - 2 (1 - x) p_a , \nonumber \\
& f_1 = 2 \ae (\tilde a_2 p_a T_{cb} + \tilde b_2 p_b T_{ca} ) - x^2 \frac{T_{cb}}{\tilde b_2} - (1 - x)^2 \frac{T_{ca}}{\tilde a_2}, \nonumber \\
& f_0 = \ae T_{ca} T_{cb}.
\end{align}
The critical temperature of the mixture as a function of the mixing parameter $x \in [0,1]$ is shown in Fig.~\ref{fig1}
The parameters of interaction used for plotting is given in Appendix~\ref{appA}. Note that at $x=0$ we have a single system of species $b$, for which $T_c^{(b)} = 4.8028$. The case $x=1$ means existence of a single system of species $a$, for which $T_c^{(a)} = 3.9502$.    

The reduced form of the equation \eqref{3d11} is
\begin{equation}\label{3d20}
\rho_+^3 + p_s \rho_+ + q_s = 0
\end{equation}
where
\begin{equation}\label{3d21}
p_s = - \frac{6 D_{ab}(x,T)}{a_+}, \qquad q_s = - \frac{6 M_{ab}}{a_+}. 
\end{equation}
For all $\tau > 0$ the equation \eqref{3d20} has a single solution 
\begin{equation}\label{3d22}
\rho_+ = \left(\frac{3}{a_+}\right)^{\frac13}\left( A^{\frac13} + B^{\frac13} \right),
\end{equation}
where
\begin{align}\label{3d23}
& A = M_{ab} + \sqrt{M_{ab}^2 - M_{0s} D_{ab}^3 }, \nonumber \\
& B = M_{ab} - \sqrt{M_{ab}^2 - M_{0s} D_{ab}^3 }. 
\end{align}
In these expressions
\begin{equation}\label{3d24}
M_{0s} = \frac{8}{9a_+}.
\end{equation}

We find the temperature dependence of $D_{ab}$ from \eqref{3d8}. Write an identity $D_{ab} = D_{ab} (x,T) - D_{ab} (x,T_c)$, namely according to \eqref{3d12} subtract zero from $D_{ab}$. Then
\begin{align}\label{3d25}
D_{ab}  = & \; x^2 \left( \frac{T_c}{W_a(0, T_c)} - \frac{T}{W_a(0)} \right)  +  (1-x)^2 \left( \frac{T_c}{W_b(0, T_c)} - \frac{T}{W_b(0)} \right)  .
\end{align}  
Taking into account the relation
\begin{align}\label{3d26}
& \tilde a_2 W_a(0) \! = \! T_{ca} \!  + \!  2 \tilde a_2 p_a T \! = \! \tilde a_2 W_a(0, \!  T_c) \! + \! 2 \tau \tilde a_2 p_a T_c , \nonumber \\
& \tilde b_2 W_b(0) \! = \! T_{cb}  \! +  \! 2 \tilde b_2 p_b T \! = \! \tilde b_2 W_b(0, \!  T_c) \! + \! 2 \tau \tilde b_2 p_b T_c,
\end{align}
we obtain $D_{ab}$ in more compact form
\begin{align}\label{3d27}
D_{ab}  = & \; x^2 \tau \frac{W_a(0, T_c) - 2 p_a}{ \beta_c W_a(0, T_c) \beta_c W_b (0)}  +  (1-x)^2 \tau \frac{W_b(0, T_c) - 2 p_b}{ \beta_c W_b(0, T_c) \beta_c W_a (0)}   .
\end{align}  
The coefficient $D_{ab}$ is negative for all $\tau > 0$, and positive at $T<T_c$. It is proportional to $\tau$, like a similar coefficient in the case of a single system~\cite{ref16}. 

\begin{figure}[h!]
	\centering \includegraphics[width=0.5\textwidth]{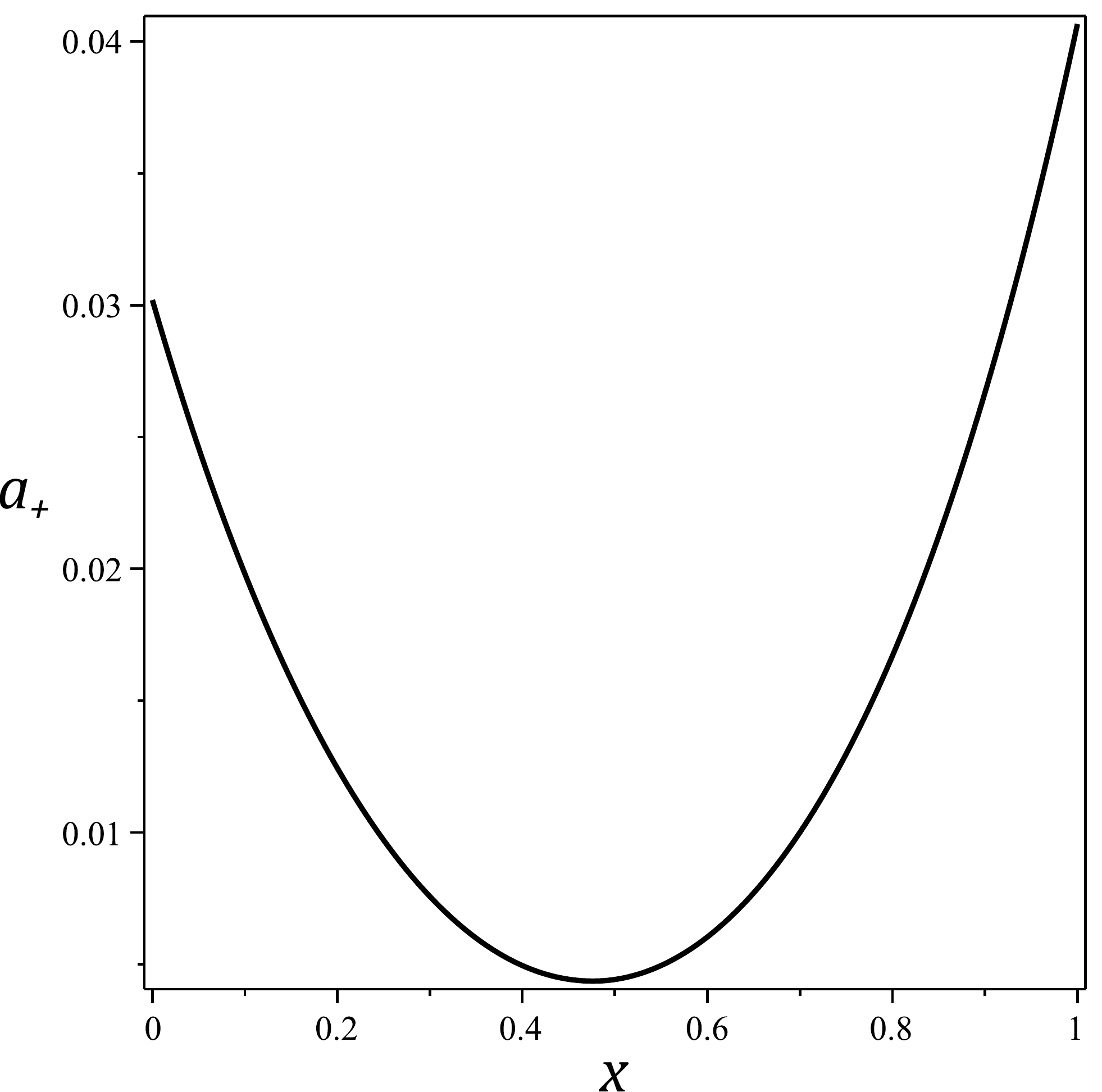}\\
	\vskip-3mm\caption{Plot of the coefficient $a_+$ given in \eqref{3d8} as a function of $x$}\label{fig2}
\end{figure}

Solutions of the equation \eqref{3d20} are similar to solutions of the analogous equation in the case of a single system (see \cite{ref16}). The difference is that the critical temperature of the mixture depends on the mixing parameter $x$, as well as the coefficient $a_+ = a_+(x) $. The latter case is shown in Fig.~\ref{fig2}.

In the temperature range $T<T_c$ the equation \eqref{3d20} has three real solutions, which we can find using the method suggested in the article~\cite{ref16}. 

\begin{figure}[h!]
	\centering \includegraphics[width=0.5\textwidth]{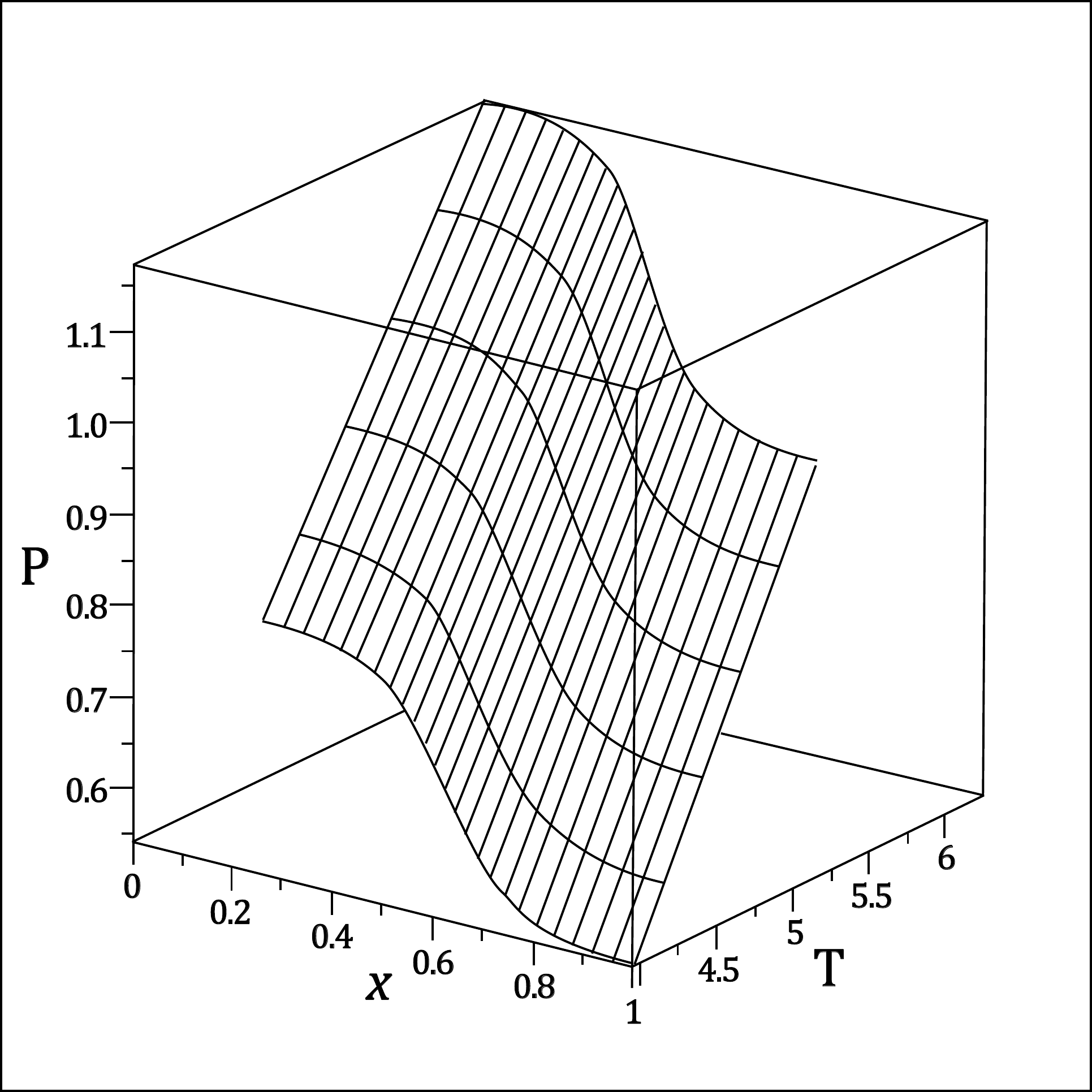}\\
	\vskip-3mm\caption{Plot of the surface of the pressure extremum.}\label{fig3}
\end{figure}

The solutions $\rho_+$ are functions of the temperature $T$, the effective chemical potential $M_{ab}$ and the mixing parameter $x$. Therefore, it is possible to visualize the function $\rho_+(T, M_{ab}, x)$ only as separate projections on the surfaces:
\begin{align}\label{3d28}
& \rho_+ \Big|_{T = const} = \rho_+^{(1)} (M_{ab}, x) , \nonumber \\
& \rho_+ \Big|_{x = const} = \rho_+^{(2)} (M_{ab}, T) , \\
& \rho_+ \Big|_{M_{ab} = const} = \rho_+^{(3)} (T, x) . \nonumber
\end{align}  
Each of these projections reflect particular physical process. $\rho_+^{(3)}$ is now important since it determines the Widom line of a binary mixture.  Easy to make sure, that for all $T > T_c$ at $M_{ab}=0$ we have $\rho_+ (M_{ab}=0) = 0$. Then find the pressure using the well-known formula
\[
PV = k_B T \ln \Xi
\] 
and obtain
\begin{equation}\label{3d29}
Pv = k_B T \left[ \frac{1}{N_v} \ln{( g_v^{(a)} g_v^{(b)} )} + E_W^{(a)} + E_W^{(b)} \right],
\end{equation}
where, according to \eqref{2d16}, read
\begin{align}\label{3d30}
& E_W^{(a)}  = a_0 - \frac{\beta}{2}( \mu_a^{(W)} )^2 + m_a ( a_1 + \mu_a^{(W)}) + \frac{m_a^2}{2} d_a(0) + \frac{a_3^4}{8 a_4^3}, \\
& E_W^{(b)}  = b_0 - \frac{\beta}{2}( \mu_b^{(W)} )^2 + m_b ( b_1 + \mu_b^{(W)}) +  \frac{m_b^2}{2} d_b(0) + \frac{b_3^4}{8 b_4^3},\nonumber 
\end{align}
moreover,
\begin{align}\label{3d31}
& \mu_a^{(W)} = -a_1 - m_a d_a(0) + \frac{a_3^3}{a_4^2}, \nonumber \\
& \mu_b^{(W)} = -b_1 - m_b d_b(0) + \frac{b_3^3}{b_4^2}
\end{align}
The temperature dependence hides in the quantities $d_\delta(0)$. Easy to see from the formulas above, that the pressure along the Widom line is a function of the relative temperature $\tau$ and the parameter $x$. The dependence on $x$ eventuates from $d_\delta(0)$ containing the critical temperature $T_c (x)$, which is a function of the mixing parameter:
\begin{align}\label{3d32}
& d_a(0) \! = \! -\tau \tilde a_2 \frac{T_c}{T_c^{(a)}} \frac{1-\omega_{0a}}{1+ \omega_{0a}\tau_a} \! - \! \tilde a_2 \frac{T_c \! - \! T_c^{(a)}}{T_c^{(a)}} \frac{1-\omega_{0a}}{1+ \omega_{0a}\tau_a}, \nonumber \\
& d_b(0) \! = \! -\tau \tilde b_2 \frac{T_c}{T_c^{(b)}} \frac{1-\omega_{0b}}{1+ \omega_{0b}\tau_b} \! - \! \tilde b_2 \frac{T_c \! - \! T_c^{(b)}}{T_c^{(b)}} \frac{1-\omega_{0b}}{1+ \omega_{0b}\tau_b},
\end{align}
where $\tau_a$, $\tau_b$ are given in \eqref{2d22}, and for $\omega_\delta$ read
\begin{equation}\label{3d33}
\omega_{0a} = 2 \tilde a_2, \quad \omega_{0b} = 2 \tilde b_2.
\end{equation}
At $x=0$ we get the Widom line of a single subsystem of species $b$, and at $x=1$ \--- of species $a$. For all $x \in (0,1)$ the set of Widom lines forms a surface represented in Fig.~\ref{fig3} 

\section{The equation for density}\label{sect_density} 

General principles of statistical mechanics give the equation for average numbers of particles of both species $<N_a>$ and $<N_b>$. Denote
\begin{equation}\label{4d1}
n_a = \frac{<N_a>}{N_v}, \quad n_b = \frac{<N_b>}{N_v},
\end{equation}
and then, according to \cite{ref14,ref16}, read
\begin{equation}\label{4d2}
n_a = \frac{1}{N_v}\frac{\partial \ln{\Xi}}{\partial \beta \mu_a}, \quad  n_b = \frac{1}{N_v}\frac{\partial \ln{\Xi}}{\partial \beta \mu_b}.
\end{equation}
Taking into account \eqref{3d1} find
\begin{align}\label{4d3}
& n_a = - M_a + n_c^{(a)} + \tilde a_2 \gamma \tau_a \rho_n, \nonumber \\
& n_b = - M_b + n_c^{(b)} + \tilde b_2 \gamma \tau_b \rho_n.
\end{align}
The quantities $M_a, M_b$ are given in \eqref{3d3}, that means they are functions of  $\tau$ and $\rho_+$. The equalities \eqref{3d3} give the following equation 
\begin{equation}\label{4d4}
x^3 \rho_n^3 + g_a x \rho_n + q_a = 0,
\end{equation}
where 
\begin{equation}\label{4d5}
g_a = \frac{6 \tilde a_2 }{a_4}, \quad q_a = \frac{6 (n_c^{(a)} - n_a)}{a_4}.
\end{equation}
From \eqref{4d3} we have
\begin{equation}\label{4d6}
(1 - x)^3 \rho_n^3 + g_b (1 - x) \rho_n + q_b = 0,
\end{equation}
here
\begin{equation}\label{4d7}
g_b = \frac{6 \tilde b_2 }{b_4}, \quad q_b = \frac{6 (n_c^{(b)} - n_b)}{b_4}.
\end{equation}

Before writing solutions of the equations \eqref{4d4} and \eqref{4d6}, we introduce the total density of the mixture $ n_+ $ and the concentration $ \eta $ using the equations
\begin{equation}\label{4d8}
n_a = \eta n_+, \quad n_b = (1-\eta) n_+.
\end{equation}
Then
\begin{equation}\label{4d9}
n_+ = n_a + n_b, \quad \eta = \frac{n_a}{n_+}, \quad 1-\eta = \frac{n_b}{n_+}.
\end{equation}
From \eqref{4d4} we have
\begin{equation}\label{4d10}
x \rho_n = -2 \bar a \cos{\left(\frac{\alpha_n + \pi}{3}\right)},
\end{equation}
where
\begin{equation}\label{4d11}
\alpha_n = \arccos{\left( \frac{n_c^{(a)} - \eta n_+}{n_c^{(a)} }\right) }, \quad \bar a = \left( - \frac{2\tilde a_2}{a_4}\right)^{\frac12} .
\end{equation}
From \eqref{4d6} we have
\begin{equation}\label{4d12}
(1 - x) \rho_n = -2 \bar b \cos{\left(\frac{\beta_n + \pi}{3}\right)},
\end{equation}
where
\begin{equation}\label{4d13}
\beta_n \! = \! \arccos{ \! \left( \! \frac{n_c^{(b)} \! - (1 - \eta) n_+}{n_c^{(b)} } \! \right) }, \;\;\; \bar b =  \! \left( - \frac{2\tilde b_2}{b_4}\right)^{\frac12}   \!\!\!  .
\end{equation}
The sum of the expressions \eqref{4d10} and \eqref{4d12} equals
\begin{equation}\label{4d14}
\rho_n = -2 \bar a \cos{\left(\frac{\alpha_n + \pi}{3}\right)} - 2 \bar b \cos{\left(\frac{\beta_n + \pi}{3}\right)}.
\end{equation}
Note that $\rho_n$ fails to be the function of temperature. It has a restricted range since $\bar a$ and $\bar b$ are finite quantities.
The subtract of \eqref{4d10} and \eqref{4d12} gives the equation
\begin{equation}\label{4d15}
x = \frac12 \left[ 1 + \frac{ \bar a \cos{\left(\frac{\alpha_n + \pi}{3}\right)} - \bar b \cos{\left(\frac{\beta_n + \pi}{3}\right)} }{ \bar a \cos{\left(\frac{\alpha_n + \pi}{3}\right)} + \bar b \cos{\left(\frac{\beta_n + \pi}{3}\right)} } \right], 
\end{equation}
which allows us to find the relationship between the mixing parameter $ x $ and the concentration $ \eta $ for an arbitrary value of the total density $ n_+ $.


\section{Conclusions}

In this article, we represent the investigation of the behavior of a binary asymmetric mixture at $T > T_c$. Among the results is a two-dimensional analog of the Widom line for different values of the mixing parameter $x \in [0,1]$.  The equation of state of the system, which includes the dependence on the parameter $ x $, is calculated. This parameter links to the concentration and is also used to determine the critical temperature of the mixture. In the boundary cases $ x = 0 $ and $ x = 1 $, the formulas obtained here describe separately subsystems of particles of species $ b $ and $ a $, respectively.  In the case of $ x \in (0,1) $, we have a mixture of components, that is characterized by the critical temperature $ T_c $ (see Fig.~\ref{fig1}). The Widom line shifts on the surface shown in Figure \ref{fig3} as the composition of the mixture changes. 
\appendix

\section{Parameters of the model}\label{appA}

The following values of the interaction potential parameters are used for numerical calculations and plots.

For both species $\alpha_a = \alpha_b = \alpha$, as well as $R_0^{(a)} = R_0^{(b)} = R_0$:
\begin{equation}\label{a1}
R_0 = 5.3678, \quad \alpha = 1.8167, \quad  \alpha_R = 0.3385,
\end{equation}
which coincide with the parameters of the Morse potential for sodium~\cite{ref20,ref21}. 
We choose the following values of the parameters $\gamma_a$ and $n_0^{(a)}$ of the potential $U_a (r)$ (see. \eqref{2d1}) corresponding to species $a$:
\begin{equation}\label{a2}
\gamma_a = 1.6500, \quad n_0^{(a)} = 1.7255.
\end{equation}
The values \eqref{a2} give
\begin{equation}\label{a3}
p_a = 1, \quad \alpha_a^* = 5,
\end{equation}
see \eqref{2d12}. This set of parameters eventuates in the following values of the coefficients $a_n$ (see. \eqref{2d11} and \eqref{2d18})
\begin{align}\label{a4}
& a_0 = 1.1220, \quad a_1 = 0.7506, \qquad \! a_2 = 0.3413, \nonumber \\
& a_3 = 0.0265, \quad a_4 = -0.0407, \quad \tilde a_2 = 0.3500. 
\end{align}

The values of parameters $\gamma_b$ and $n_0^{(b)}$ of the potential $U_b (r)$ differ from $\gamma_a$ and $n_0^{(a)}$
\begin{equation}\label{a5}
\gamma_b = 1.6000, \quad n_0^{(b)} = 1.7613.
\end{equation}
They give 
\begin{equation}\label{a6}
p_b = 1, \quad \alpha_b^* = 5.5,
\end{equation}
and the following values of the coefficients $b_n$ 
\begin{align}\label{a7}
& b_0 = 1.1951, \quad b_1 = 0.7832, \qquad \! b_2 = 0.3437, \nonumber \\
& b_3 = 0.0231, \quad b_4 = -0.0302, \quad \tilde b_2 = 0.3526. 
\end{align}
The normalization constants given in \eqref{2d2} equal
\begin{align}\label{a8}
& C_H^{(a)} = 1.2544, \quad A_a = 0.2217, \nonumber \\
& C_H^{(b)} = 1.2938, \quad A_b = 0.2742. 
\end{align}
The values of the critical temperatures of separate single systems of species $a$ and species $b$ are
\begin{equation}\label{a9}
k_B T_c^{(a)} = 3.9502, \qquad k_B T_c^{(b)} = 4.8028.
\end{equation}

\end{document}